\documentstyle[twocolumn,prl,aps,graphicx]{revtex} 

\begin{document} 
\draft 
\title{ Phase-Controlled Force and Magnetization Oscillations\\ 
 in Superconducting Ballistic Nanowires.} 
\author{I.V.Krive$^{1,2}$, I.A.Romanovsky$^{3}$, 
E.N.Bogachek$^{3}$, and Uzi Landman$^{3}$} 
\date{\today} 
\address{$^{1}$Department of Applied Physics, Chalmers University of 
Technology and G\"{o}teborg University, SE-412 96 G\"{o}teborg, 
Sweden \\ 
$^{2}$B. I. Verkin Institute for Low Temperature Physics and 
Engineering, Lenin ave. 47, Kharkov 61103, Ukraine\\ 
$^{3}$ School of Physics, Georgia Institute of Technology,
 Atlanta, Georgia 30332-0430} 
 
\maketitle 
\begin{abstract} 
{ 
The emergence of 
superconductivity-induced phase-controlled forces in the
 ($10^{-2}-10^{-1}$) nN range, and of magnetization oscillations, 
in nanowire junctions, is discussed.
 A giant magnetic response to applied weak magnetic fields, 
 is predicted in the ballistic Josephson junction formed by
 a superconducting tip and a surface, bridged by a normal metal nanowire
where Andreev states form.} 
\end{abstract} 

\narrowtext 
 
Formation, energetics and mechanical properties of metallic nanowires (NWs), 
 were predicted in early simulations [1], 
and they have been the subject 
of subsequent significant research
 endeavors [2].
For normal metals 
the oscillatory behavior of the elongation forces [1] with the size of a NW
formed between a surface and a retracting tip
 have been
 shown to be correlated with a quantized staircase
 behavior of the electrical conductance [3-7]. 
 However, the influence of superconductivity (SC) on the nanomechanical
 properties of such NWs has not been explored yet, and these effects are the
 focus of this Letter. 
 
In normal metals 
the cohesive force in an atomic-scale metallic contact 
can be estimated as 
$F_n\sim\varepsilon_F/\lambda_F$ ,  where $\varepsilon_F$ and 
$\lambda_F$ are the Fermi energy and wavelength.
 The onset of SC introduces 
a new energy scale, i.e. the superconducting gap $\Delta\ll\varepsilon_F,$
 and a new lengthscale, i.e. the SC coherence length 
 $\xi_0=\hbar v_F/\pi\Delta\ .$ 
On first sight, the resulting SC- 
induced forces are expected to be of the order of 
$\ F_{sc}\sim\Delta/\xi_0$, and when added to the aforementioned 
normal-metal forces ($F_n$ , which are of the order of several nN ),
they are estimated to be below the  
atomic force microscopy (AFM) detection limit [8].
However, in the superconducting regime, 
under certain conditions all the transverse channels ($N_{\perp}$) supported 
by the junction will contribute 
coherently to the free energy, and when $N_{\perp}\gg 1\;$ 
the above consideration may result in a gross underestimation
 of $F_{sc}$. 
In particular, we predict that under 
favorable conditions, in a nanowire connecting two superconducting 
electrodes, modelled here as a transparent short ($L\ll\xi_0$) 
superconducting-normal-superconducting (SNS) junction, a (measurable) force
 \mbox{$\;F_{sc}\sim (L/\xi_0)(\varepsilon_F/\lambda_F)$} would be manifested -
 we refer to these forces as Andreev forces (AF); 
in long SNS junctions the 
AF are expected to be significantly smaller.

To calculate the SC - induced contribution
 to the total elongation force we model the junction comprised of the
 superconducting tip and surface and of the NW bridging
 them, as an SNS junction
(the NW remains normal due to the suppression of superconductivity
 in very small structures [9]).
 Andreev reflections occur at the SN boundaries of the junction,
 and the dimensions of the NW are such that transport through it is ballistic. 
This holds true for short SNS junctions, while in long ones
 impurity scattering may occur, resulting in a reduced transmission probability.
 
The force equals the spatial derivative of the grand-canonical
potential $\Omega$, i.e. \ $ F=-{\partial\Omega}/{\partial L} $. 
In superconducting junctions there are two contributions to the 
AF, $ F_{A}=F_{\perp}+F_L $ ;\ 
$F_{\perp}$ is related to the dependence of the number of quantized 
transverse channels, $N_{\perp}$, on the degree of elongation, and $F_{L}$ 
originates from the dependence of the Andreev bound states on the length
 of the junction.

{\it Short ($\,L\ll\xi_0$) junctions.} 
In the following we focus on the SC-induced force oscillations in short
 junctions. 
Since here the Andreev states are independent of the mode index
 ${[10]}$, the potential can be written as 
$\Omega_s(\varphi,L)\approx N_{\perp}(L)\Omega_A(\varphi,L) 
$, 
where $\Omega_A(\varphi,L)$ corresponds to a single-channel SNS junction. 
In a cylindrical geometry 
$ 
N_{\perp}(L)\simeq{\pi V}/{\lambda^2_FL}, 
$ 
where $V$ is the volume ( which remains constant during the elongation process
 ${[1]}$ ).

 In general, both 
bound  (superscript (b) below) and scattering Andreev states contribute 
to $\Omega_A(\varphi,L)$; for our purpose 
only the bound states are important (see below), and thus 
$ 
F^{(b)}_{A}=F^{(b)}_{\perp}+F_{L}^{(b)} 
$ 
,where 
 
\begin{eqnarray} 
F^{(b)}_{\perp}=\frac{N_{\perp}}{L}\Omega^{(b)}_{A}(\varphi,L),\qquad 
F_{L}^{(b)}=-N_{\perp}\frac{\partial\Omega^{(b)}_{A}(\varphi,L)}{\partial L}. 
\end{eqnarray} 
The spectrum of Andreev bound states in a single-mode ballistic SNS junction of
 length $L$ and transparency $D\leq 1$ is found from
 the equation (see e.g. Ref.[11]) 
\begin{equation} 
\cos\left(2\arccos\frac{E}{\Delta}-2\frac{E}{\Delta_L}\right)=R+D\cos\varphi\ , 
\end{equation} 
where $\Delta_L\equiv\hbar v_F/L ,\ \ D+R=1$, and $\varphi$ is the
 superconducting phase difference. The distinction between different channels
 enters through $\Delta_L$, and it is neglected for short junctions.
 A reduced NW transparency reflects
 the presence of interfacial barriers, such as those likely to
 form at tip-surface contacts.
 From Eq.(2) we obtain in the limit  $L\ll\xi_0$ the bound-state energies

\begin{equation} 
\!\!\!\!\!\! E^{(+)}_0\!=
\!-\! E^{(-)}_0\!\!\simeq
\Delta W(\varphi)
\left(
1-\frac{L\sqrt{D}}{\xi_0}
\left\vert
\sin\frac{\varphi}{2}
\right\vert
\right)\ , 
\end{equation} 

\noindent where $W(\varphi)=\sqrt{1-D\sin^2\frac{\varphi}{2}}$ . 
Using Eq.(1), with $\Omega_A$ taken simply as the thermodynamic potential
 for a two-level 
($E^{(+)}_0, E^{(-)}_0$)  system, yields  
\begin{equation} 
F^{(b)}_{\perp}\simeq-2N_{\perp}\frac{T}{L} 
\ln\left\{ 
\frac 
{\cosh^2\left( 
{W(\varphi)\Delta 
}/ 
{2T} 
\right)} 
{\cosh^2({\Delta}/{2T})} 
\right\}, 
\end{equation} 
\begin{eqnarray} 
F^{(b)}_L \simeq 
2N_{\perp}\frac{\Delta}{\xi_0} 
W(\varphi) 
\sqrt{D\sin^2\frac{\varphi}{2}} 
\tanh\left( 
\frac{\Delta}{2T}W(\varphi) 
\right). 
\end{eqnarray}

\indent For low transparency junctions $(D\ll1)$ 
$F^{(b)}_{\perp}$ may be approximated as 
\begin{equation} 
F^{(b)}_{\perp}\simeq\frac{N_{\perp}\Delta}{L}D\sin^2{\frac{\varphi}{2}} 
\tanh\left(\frac{\Delta}{2T}\right) 
\sim\frac{N_{\perp}\Delta}{L}D \ , 
\end{equation} 
and since $F^{(b)}_{L} \sim N_{\perp}(\Delta/\xi_0)\sqrt{D}$ it is evident that 
$\vert F^{(b)}_{\perp}\vert\gg\vert F^{(b)}_L\vert $, provided that
$D\gg(L/\xi_0)^2$. 
In contrast, in low transparency junctions 
$D\ll(L/\xi_0)^2\ll 1 $ 
and the $F^{(b)}_L $ 
contribution dominates.\\ 
\indent 
\indent For point contacts (i.e. extremely short junctions) the AF can be
 calculated (when $D\gg(L/\xi_0)^2$ ) by taking the limit 
$\,L/\xi_0\rightarrow 0$ for the Andreev bound states (Eq.(3)). 
In this case, continuum states do not affect the free energy 
(see e.g. Ref.[11]) and, therefore, they do not contribute to the force. 
 In contrast, for low transparency junctions the $\,L/\xi_0$-corrections 
to the bound state energies determine the force oscillations. 
To this order the continuum states do contribute to the free energy and they
 can change the dependence of the AF on the phase difference. 
Thus, Eq.(5) can be considered as an estimate of the AF in
 a short junction. 
the contribution of the continuum states ($F^{(c)}_A$)
is extremely small, i.e. $\ F^{(c)}_A(D\ll 1)\ll(L/\xi_0)^3 
(\varepsilon_F/\lambda_F)$. 
 
The phase-dependent force in a superconducting quantum point contact (QPC) 
($D=1$) is related to the quantized Josephson current 
 $J_s$ (see Ref.[10]) 
\begin{equation} 
J_s = \frac{e}{\hbar}\left(-L\frac{\partial F^{(b)}_A}{\partial\varphi} 
\right) = N_{\perp}\frac{e\Delta}{\hbar}\sin\frac{\varphi}{2}. 
\end{equation} 
The force oscillations (portrayed by the dependence of the force on the 
contact size) are determined by two distinct contributions: (i) a large 
phase-independent term ( operative also in normal-metal NWs )
of the order of $\,N_{\perp}\varepsilon_F/ 
\lambda_F$ originating from incoherent contributions of all the conducting 
electrons to the thermodynamic potential [4-6], and (ii) a coherent
 SC-induced force (Eq.(4)-(6)). It is the latter, phase-dependent,
 term that is 
directly related to the quantized Josephson current.

The amplitude of the AF oscillations  may be readily estimated as follows:
 for 
$D\sim 1$ and $L\ll\xi_0$ the amplitude of the Andreev force is 
of the order of 
$\ F^{(b)}_{A}\sim N_{\perp}\Delta/L \sim
 L \varepsilon_F/\xi_0\lambda_F\sim(L/\xi_0)\left[nN\right]$;
 in the ballistic regime
 for a non-transition metal $\xi_0\sim 10^{-5}-10^{-4}cm $.
Using state-of-the-art instrumentation such forces
 (e.g. $10^{-2} - 10^{-1}nN$),
 can be measured ${[8]}$. 

{\it Long $(L\gg\xi_0)$ junctions.} 
For a long transparent \mbox{$(D=1)$} junction 
the spectrum of the Andreev-Kulik (AK) levels $|E_n|\ll\Delta$ 
takes the form ${[12]}$ 
\begin{equation} 
E_n^{\pm}\simeq
\pi\Delta_L\left(\pm\frac{\varphi}{2\pi}+\frac{1}{2}+n\right),
 \,\ n=0,\pm 1,\pm 2,\dots 
\end{equation} 
Since this spectrum does not depend on the superconducting gap $\Delta$, 
for temperatures $T\ll\Delta$ all the thermodynamic 
properties of a long ballistic junction are essentially 
material independent. Evaluation of the Josephson current in this case is
 equivalent to the calculation of the 
persistent current for chiral fermions on a ring 
with circumference $2L$ [13]. 
The corresponding phase-dependent part of the thermodynamic potential
 $\Omega_{A}(\varphi)$ can be evaluated for the AK spectrum, yielding

\begin{equation} 
\tilde\Omega_A(\varphi)=4T\sum\limits_{k=1}^{\infty}\frac{(-1)^k}{k} 
\frac{\cos k\varphi}{\sinh({2\pi T k}/{\Delta_L})}\ . 
\end{equation} 

 The force oscillations induced by the AK level structure 
 in a single-channel long SNS junction are ($\vert\varphi\vert\leq\pi$) 
\begin{equation} 
\tilde F_A\simeq\left\{ 
\begin{array}{cc} 
\frac{\Delta_L}{2\pi L}\left({\varphi}^2 
-\frac{\pi^2}{3}\right) 
\ ,\ \ \ T\ll\Delta_L\ \\ 
\\ 
-16\pi\frac{T^2}{L\Delta_L}\exp\left({-\frac{2\pi T}{\Delta_L}}\right) 
\cos\varphi\ ,\ \ T\geq\Delta_L\ . 
\end{array}\right. 
\end{equation} 
We note that the above is 
equivalent to the Casimir force ${[14]}$. 
We focus here only on the phase-dependent part of the
thermodynamic potential, $\tilde\Omega_A(\varphi)$, 
and the resulting Andreev (or Casimir) force $\tilde F_A$ ,
since, as aforementioned, the force 
in superconducting junctions (Eq.(10)) is added to a much larger
phase-independent term ($\sim\varepsilon_F/\lambda_F$)
that dominates the cohesive force in metallic NWs.

In a multi-channel junction the thermodynamic potential 
is the sum over transverse channels $(ln)$ 

\begin{equation} 
\tilde\Omega=\sum\limits_{ln}\tilde\Omega_{A}^{(ln)} 
(\varphi), 
\end{equation} 
where $\tilde\Omega_{A}^{(ln)}(\varphi) $ is given by 
Eq.(9) with $\Delta_L^{(ln)}=\hbar v_F^{(ln)}/L$ 
substituted for $\Delta_L$. 
For a long junction the Fermi velocity enters explicitly 
the expression for a single channel supercurrent, and the total 
current in a multi-channel junction strongly depends on the 
junction geometry ${[15]}$. 
 
We will model the normal part of a long SNS junction by 
a cylinder of length $L$ and cross-section area $S=V/L$. 
Assuming hard-wall boundary conditions,
the electron longitudinal velocity in the $(ln)-$th channel is 
\begin{equation} 
v_F^{(ln)}(L)=\sqrt {2\left(\varepsilon_F- a\gamma^2_{ln}\right)/m}, 
\end{equation} 
 
\noindent where
 $\gamma_{ln}\ $ are the Bessel function zeroes:$J_l(\gamma_{ln})=0$, 
and $a={\hbar^2\pi L}/{2mV} $. 
The dependencies of the AF on the phase difference ($\varphi$) 
and on the length (L) of the NW are displayed, respectively, in Figs.1 and 2,
where we show $\Delta F(\varphi)=F_A(\varphi)-F_A(0)$. 
From Fig.1 we observe that the force is enhanced at special values of the phase 
difference $\,\varphi_r=\pi(2r+1),\;(r=0,\pm 1,\pm 2,...).$ 
At $\varphi=\varphi_r$ one of the AK bound states coincides 
with the Fermi energy and, most significantly, 
this state is $4N_{\perp}$-fold degenerate ${[16]}$, thus amplifying its
 contribution. 
Direct observation of the SC-induced nanomechanical effect predicted here may
 be obtained through : (i) generation of a NW of length $L$ via separation
of an AFM tip-surface contact, using a superconducting material (e.g. Pb)
at $ T < T_c $, followed by (ii)  measurement of the force required to 
maintain the NW length ($L$) as a function of variations of the phase-difference
across the SNS junction (as seen from Fig.1 this force maximizes at 
$ \varphi=\pi $ ).

 The variation of the elongation force (for $\varphi=\pi$) with the NW length
 is shown in Fig.2. We note first that even though
 the number of open channels is very large for the NW junction shown in Fig.2,
 the magnitude of the forces is significantly smaller than in the case of short
 junctions (see previous subsection)[17]. The aperiodic variations of the AF 
 originating from the change in the number of open channels upon elongation,
 are particularly pronounced at lower temperatures. 
 Note however, that such aperiodic variations occur also for normal metal
 NW [4,6] and consequently separation of the SC-induced contribution may be
 difficult.

Next we consider a multichannel 
superconducting long junction in a weak magnetic field $\mu_B B\ll\Delta\,
$ (where $\mu_B$ is the Bohr magneton),applied locally (i.e. only to the normal 
metal nanowire part of the SNS junction).
Here, the only [18] influence on the AK levels is 
through the Zeeman coupling of the electron spin ${\bf s}$ to the magnetic 
field, $H_Z=g\mu_{B}{\bf s}\!\cdot\!{\bf B}\;(g$ is the g-factor). 
The thermodynamic potential
 $\delta\Omega_A(\varphi,B)\equiv\Omega_A(\varphi,B)-\Omega_A(0,0)$
 takes the form (see. Eq.(9)) 
\begin{eqnarray} 
\delta\Omega_A(\varphi, B)=
-4T\sum\limits_{k=1}^{\infty}
\frac{(-1)^k}{k}
\frac
{(1-\cos k\varphi\cos k\chi)}
{\sinh\left({2\pi k T }/{\Delta_L}\right)}, 
\end{eqnarray} 
where $\chi\equiv \Delta_Z/\Delta_L\;$ and $\Delta_Z=g\mu_B B$ is 
the Zeeman energy splitting. 
Note that the influence of the Zeeman interaction on the thermodynamics of 
the SNS junction is equivalent to the influence of a gate voltage on the
 thermodynamic properties of quantum rings ${[19]}$. 
 
The magnetization $M_A=-\partial\delta\Omega_A(\varphi,B)/\partial B$
 at low ($T\ll\Delta_L$) and high ($T>\Delta_L$) 
temperatures is given for a single channel junction as 
\begin{eqnarray} 
M_A\simeq 
\left\{ 
\begin{array}{c} 
\!\!\!-g\mu_B\left(\frac{\chi}{\pi }\right)\ ,\ \ \ 
T=0\ \ ;\ \ \left\vert \chi \right\vert\leq\pi\\ 
\\ 
\!\!\!-8g\mu_B\frac{T}{\Delta_L}e^{-\frac{2\pi T}{\Delta_L}}
\cos\varphi\sin\left( \chi \right) , \ T\geq\Delta_L\ . 
\end{array} 
\right. 
\end{eqnarray} 
Note, that the SC-induced magnetization $M_A$, 
can be of the order of several $\mu_B$ (if $g\gg 1$) 
even for a single-channel junction, and it 
is insensitive to the superconducting phase difference at 
low temperatures. For a multichannel quantum junction 
at low temperatures the dependence of $M_A(\varphi)$ exhibits typical
 resonant behavior at the resonant phases $\varphi_r$, as shown in Fig.3. 
This is a manifestation of the effect of "giant oscillations", 
known previously for conductance oscillations ${[18]}$. 
At these phases Andreev states of 
energies $E_A=\pm g\mu_B sB$ become $2N_{\perp}$-fold degenerate ${[20]}$,
leading to giant enhancement of thermodynamic and kinetic characteristics 
of ballistic junctions in magnetic fields.
 
Since at resonance the coherent contribution
 ($\propto N_{\perp}$) of all transverse modes dominates the magnetization,
 we predict at low temperatures ($T\ll\Delta_Z$): 
(i) a giant response ($\propto N_{\perp}$) of an SNS junction
 to a magnetic field, and (ii) a step-like behaviour of the magnetization 
 as a function of the wire diameter. 
 At other values of the phase difference, 
 different transverse channels contribute to $\delta\Omega_A$ with 
different periods (i.e. in general, incoherently), 
resulting in a complex structure of the magnetic oscillations. 

In most cases a supercurrent is suppressed by the Zeeman 
interaction ${[21]}$. 
A magnetic field would also suppress the predicted Andreev force
$\delta F_A(\varphi,B)=-\partial\delta\Omega_A(\varphi,B)/\partial L$.
At low temperatures ($T\ll\Delta_L$) the force (which is periodic 
both in the phase, $\varphi$, and in the dimensionless Zeeman energy
splitting $\chi=\Delta_Z/\Delta_L$) 
can be written for a single-channel junction as 
$(\ \vert\varphi\vert ,\left\vert\chi\right\vert \leq\pi\ \ )$: 
$ 
\delta F_A\simeq 
({\Delta_L}/{2\pi L})\left[\left({\varphi}\right)^2- 
\left({\chi}\right)^2\right] . 
$ 

In summary,
 we predicted and illustrated that superconductivity induces in quantum wires 
phase-dependent forces correlated with the supercurrent. At 
resonance values of the superconducting phase difference these
 Andreev forces become measurable (nN scale). Furthermore,
 we predict giant magnetization (of the order of $N_{\perp}\mu_B$) of ballistic
 SNS junctions in a weak magnetic field at low temperatures $T\ll\Delta_Z$.
 Since low-temperature STM with superconducting tips has been already
 demonstrated ${[22,23]}$ and used to form Josephson junction ${[24,25]}$,
 and in light of improved force-detection capabilities
 (extending to $10^{-1}-10^{-3}nN$)[8],
 the above predictions provide the impetus
 for future experiments. 
 
I.V.K. acknowledges discussions with A.Kadigrobov and 
V.Shumeiko, support from the Royal Swedish Academy of 
Sciences (KVA), 
 the hospitality 
of the Georgia Institute of Technology and Chalmers 
University of Technology. 
The research of I.A.R., E.N.B., and U.L. 
was supported by the U.S. Department of Energy,
 grant No. FG05-86ER 45234, and NSF grant DMR-0205328.

\begin{figure} 
\begin{center}

\includegraphics[width=0.4\textwidth]{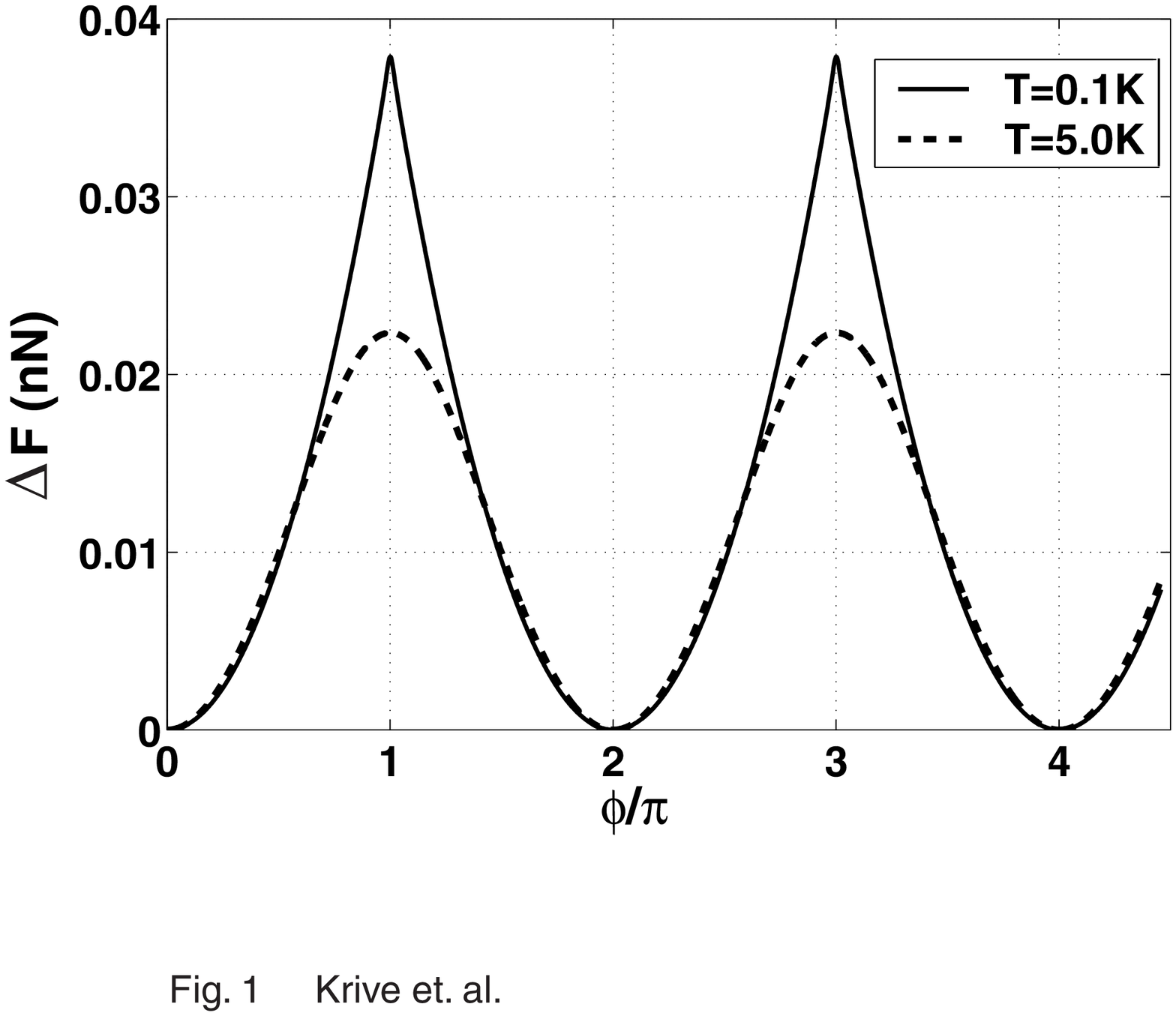}
\caption{
$\Delta F(\varphi)=F_A(\varphi)-F_A(0)$ 
vs the phase difference for a NW SNS junction. 
We use the physical parameters for Pb, i.e. 
the Fermi energy is $\varepsilon_F=1.5\times 10^{-11}erg$,
 and $v_F=1.83\times 10^{8}cm/s $.
 The volume $V=5\times 10^{-15}cm^3$,
 and the length of the junction $L=10^{-4}cm$.
 Results are shown for two temperatures, both below $T_c(Pb)=7.18K$.\ 
The force was calculated as follows: 
first, using Eq.(9), the grand canonical potential
 was found for each transverse mode 
(with different values of
 $v_F^{ln}$ and, therefore,
 different $\Delta^{ln}_L=\hbar v^{ln}_F/L$, see Eq.(12)). 
The total potential of the junction is the sum of
 contributions of all the transverse modes (Eq.(11)). 
 The derivative of the potential with respect to the length of the
 junction was evaluated numerically.} 
\label{Fig.1} 
\end{center}
\end{figure}

\begin{figure}
\begin{center} 
\includegraphics[width=0.4\textwidth]{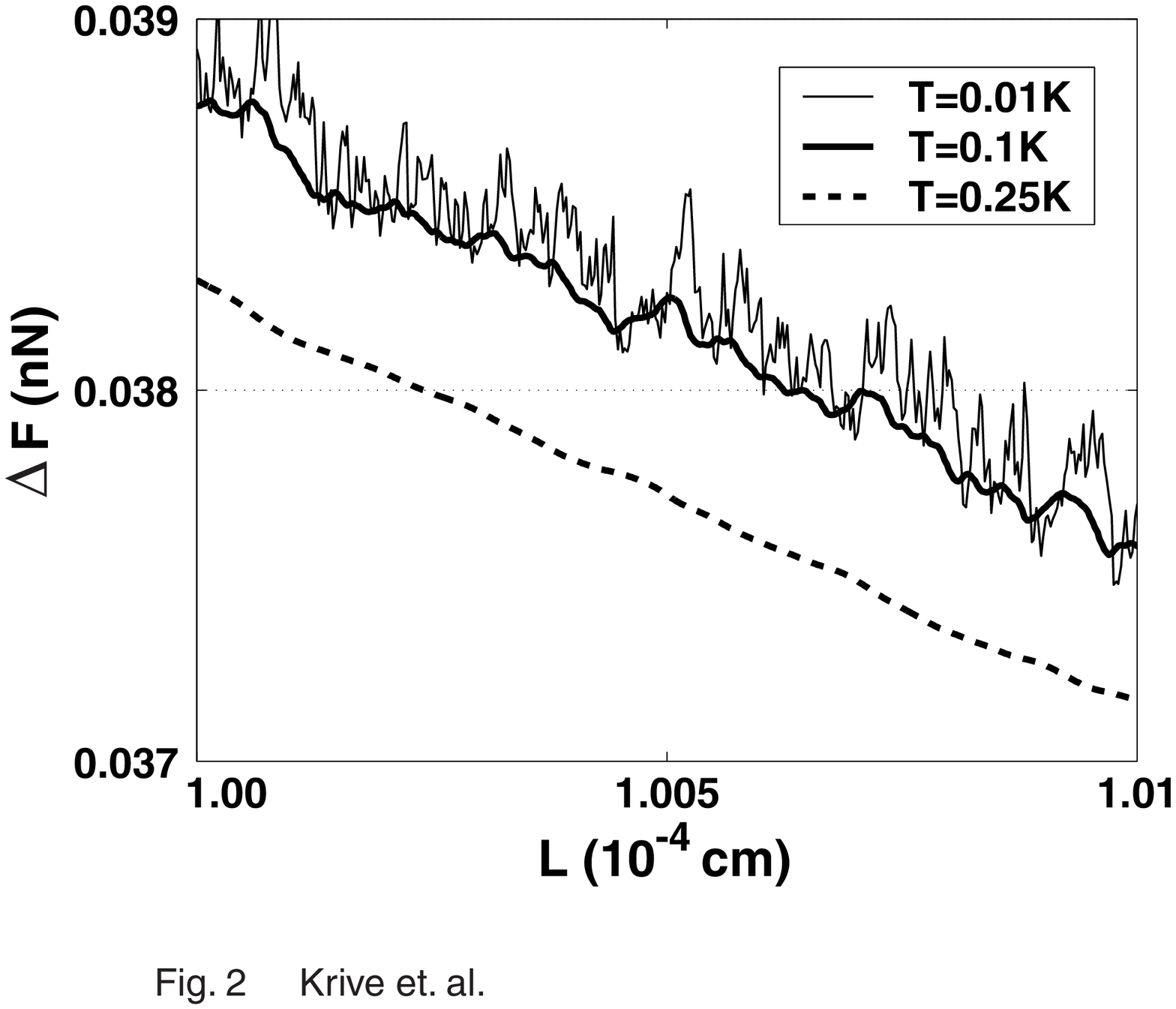}
\caption{ $\Delta F(\varphi=\pi)$
 vs the length of the junction, calculated for different 
temperatures.
 The parameters of the junction and the method of calculation are as in Fig.1.}
\label{Fig.2} 
\end{center}
\end{figure}

\begin{figure} 
%
\includegraphics[width=0.4\textwidth]{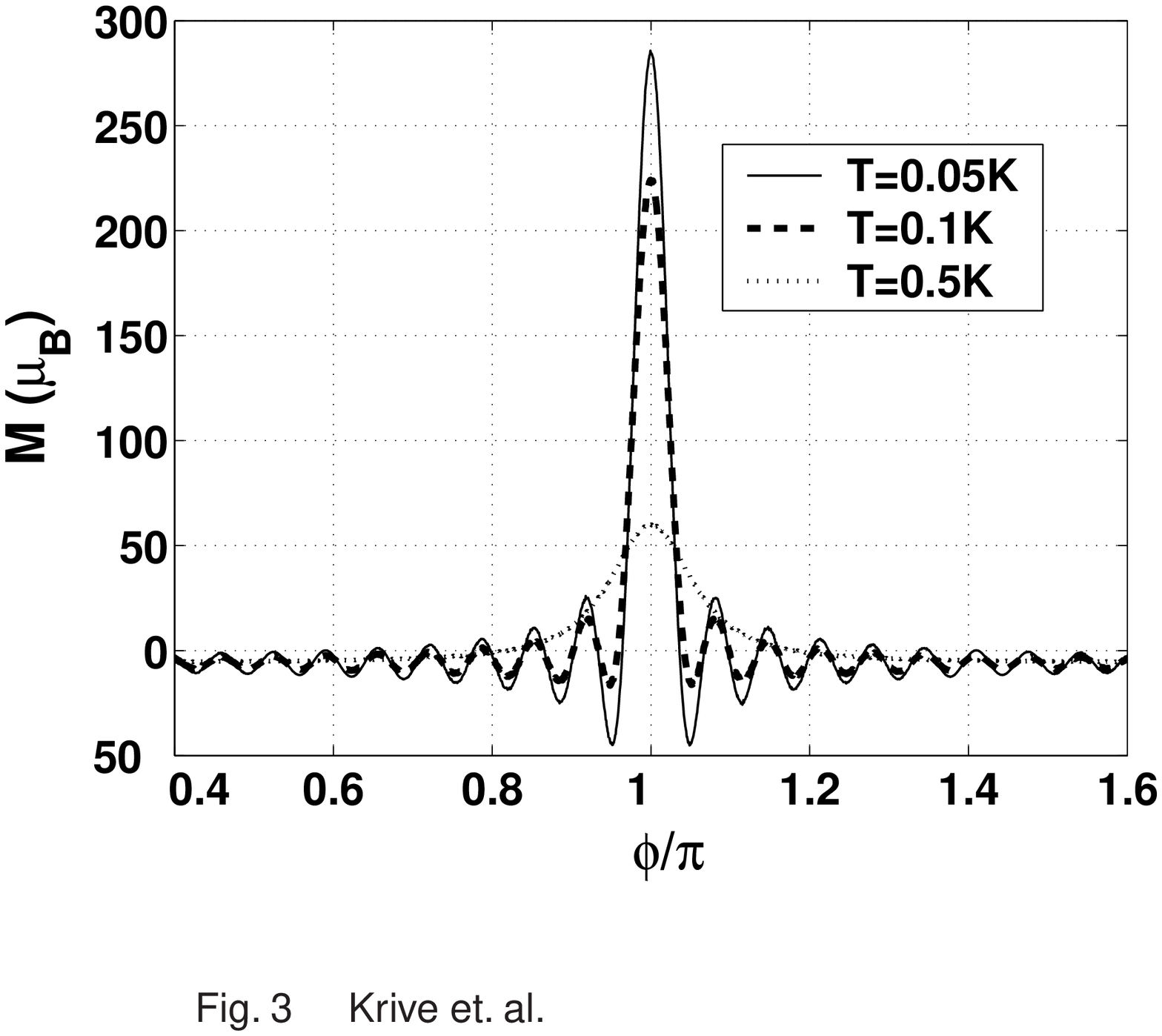}
\caption{Magnetization 
 of the junction in a 
 magnetic field of $5\ Oe$, plotted vs the phase difference $\phi$ for several
 temperatures. The parameters of the junction are as in Fig.1.
 The magnetization was evaluated as a numerical derivative of
 $\delta \Omega (\varphi, B)$\ (see Eq.(13)) with respect to B.}  
\label{Fig.3} 
\end{figure}


\begin{references}
\bibitem{[17]} For a long junction $(L\gg\xi_0)$ with only a very small number of channels($N_{\perp}\sim 1$), the force is negligibly small $\;\tilde F_A\sim \hbar v_F/L^2\sim (\varepsilon_F/\lambda_F)(\lambda_F/L)^2$. 
On the other hand,
for a relatively short ( i.e. $L\sim 0.1\xi_0\sim 10$ nm) 
ballistic NW, the SC-induced contribution to the
force can become quite large. 
For such a NW with  
 a  diameter $2R=L$ we estimate ( using the Sharvin expression 
$k_F^2/4\pi$, where $k_F(Pb)\sim 10^8 $ cm$^{-1}$ ) 
that the number of channels is $\sim$ 1500. From Eqs. (4) and (5) we
obtain an AF (evaluated at $\varphi=\pi$) of $10^{-1}$nN for $T=0.1$K, and
half this force for $T=6$K.
\end{references}
\end{document}